\documentclass[aps,prl,amsmath,amssymb,twocolumn,nofootinbib]{revtex4-1}
\pdfoutput=1
\usepackage{graphicx}
\usepackage{multirow}
\usepackage{color}
\newcommand{\beq}{\begin{eqnarray}}
\newcommand{\eeq}{\end{eqnarray}}

\newcommand{\bmp}{\noindent\begin{minipage}{16cm}}
\newcommand{\emp}{\end{minipage}\vskip 7mm} 

\usepackage{dcolumn}
\usepackage{bm}
\usepackage{bbm}
\usepackage{pxfonts}

\definecolor{rossoCP3}{cmyk}{0,.88,.77,.40}

\def\lsim{\mathrel{\rlap{\lower4pt\hbox{\hskip1pt$\sim$}}
    \raise1pt\hbox{$<$}}}                
\def\gsim{\mathrel{\rlap{\lower4pt\hbox{\hskip1pt$\sim$}}
    \raise1pt\hbox{$>$}}}                

\newcommand{\be}{\begin{eqnarray}}
\newcommand{\ee}{\end{eqnarray}}

\begin{document}
\title{\Large  \color{rossoCP3}   Jumping Dynamics}
 \author{Francesco {\sc Sannino}$^{\color{rossoCP3}{\varheartsuit}}$}
\email{sannino@cp3.dias.sdu.dk} 
\affiliation{
$^{\color{rossoCP3}{\varheartsuit}}${ CP}$^{ \bf 3}${-Origins} and the Danish Institute for Advanced Study,
 University of Southern Denmark,  Campusvej 55, DK-5230 Odense M, Denmark.}
\begin{abstract}
We propose an alternative paradigm to the conjectured Miransky scaling potentially underlying the physics describing the transition from the conformally broken to the conformally restored phase when tuning certain parameters such as the number of flavors in gauge theories.  According to the new paradigm the physical scale and henceforth also the massive spectrum of the theory jump at the lower boundary of the conformal window. In particular we propose that a theory can suddenly jump from a Quantum Chromodynamics type spectrum, at the lower boundary of the conformal window, to a conformal one without particle interpretation. The jumping scenario, therefore, does not support a near-conformal dynamics of walking type. We will also discuss the impact of jumping dynamics on the construction of models of dynamical electroweak symmetry breaking.  
\\[.1cm]
{\footnotesize  \it Preprint: CP$^3$-Origins-2012-12 \& DIAS-2012-13}
\end{abstract}

\maketitle
 
Understanding the non-perturbative dynamics of gauge theories of fundamental interactions constitutes a formidable challenge. Recently a considerable effort has been made to unveil the long distance conformal dynamics of  these theories, see  \cite{Sannino:2009za} for a recent review. 

However, despite much efforts,  we do not yet know the physical properties at the boundary between a conformally broken and a conformally restored phase for generic gauge theories. This problem remains an important mystery to  solve. A famous conjecture has been put forward some time ago \cite{Miransky:1988gk,Miransky:1996pd} and it is known as Miransky scaling. 
Here we will argue for the potential existence of another intriguing possibility leading to a radically different near-conformal behavior. For reader's convenience and for setting up the stage we start with a brief, but self-contained, review of the the Miransky scaling and modeling. We will then introduce the alternative scenario and deduce the consequences for models of dynamical electroweak symmetry breaking.
 
 \section*{Miransky scaling and walking dynamics}
This  scaling arises under the following assumptions: i) A given gauge theory possesses simultaneously, at least, a non-trivial infrared (IR) fixed point and an ultraviolet (UV) one; ii)  Upon changing an external parameter of the theory, e.g. the number of flavors, at a critical value of this parameter the IR fixed point merges with the UV fixed point; iii) This merging is sufficiently smooth that the nearby conformal phase is felt, in the conformally broken phase, for values of the external parameter near the phase transition. 

\begin{figure}
\begin{center}
\includegraphics[width=8truecm]{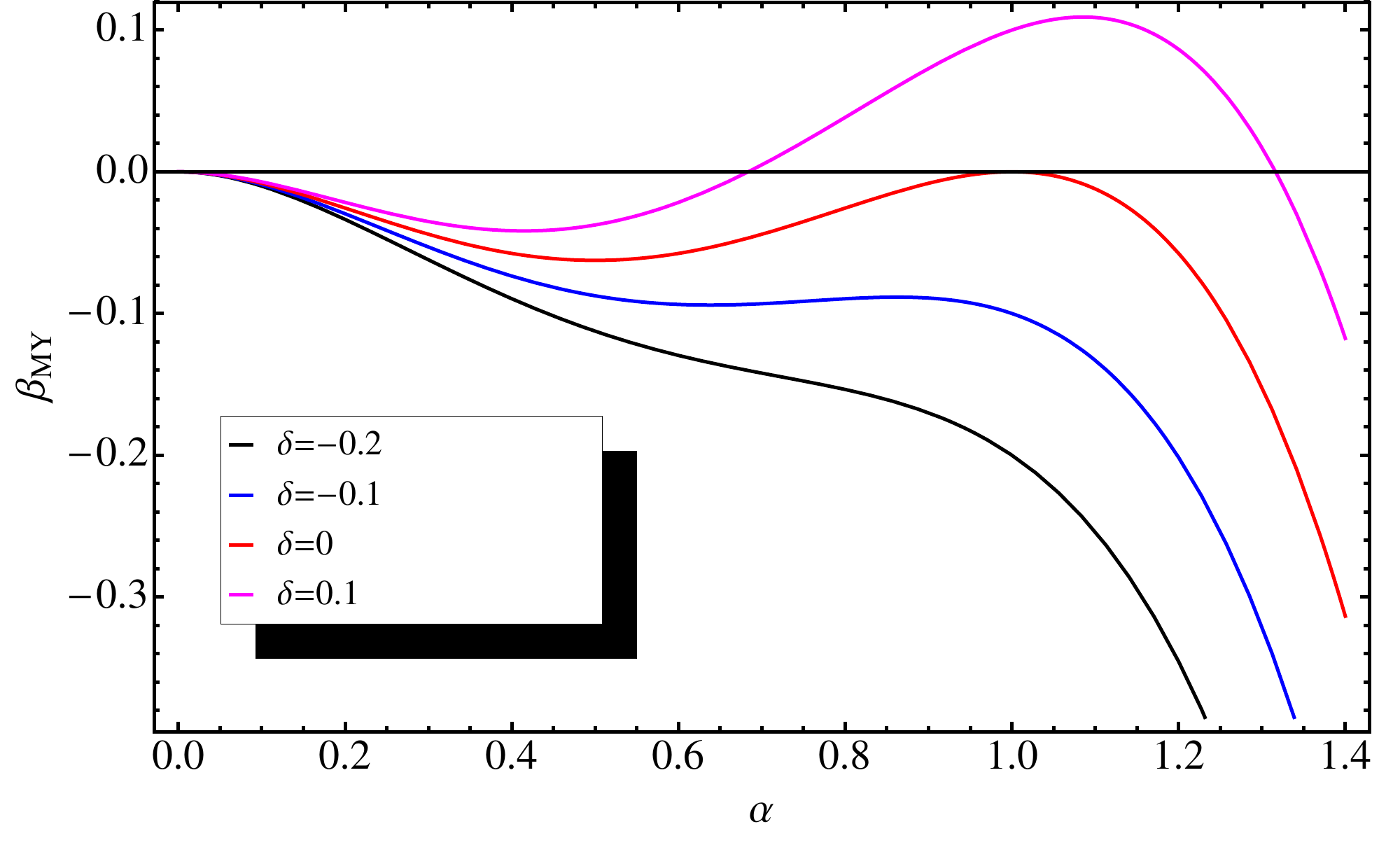}
\caption{$\beta_{MY}$ for different values of $\delta = n_f - n_f^c$.} \label{Beta_Miransky}
\end{center}
\end{figure}

Without loss of generality it is possible to model the beta function near the critical number of flavors as follows: 
\begin{equation}
\beta_{MY}= - {\alpha^2}\left({\alpha -1 -\sqrt{\delta}}\right)\left( {\alpha-1+\sqrt{\delta}}\right) = -\alpha^2 ((\alpha -1)^2 - \delta)\ .
\end{equation}
The double zero at the origin embodies asymptotic freedom and  
$\delta = n_f - n_f^c$.
For positive values of $\delta$ the beta function possesses a non-trivial IR and  UV fixed point at the following values of the coupling:
\begin{equation}
\alpha_{IR} = 1 -\sqrt{\delta} \ , \qquad {\rm and} \qquad \alpha_{UV}=1+\sqrt{\delta} \ .
\end{equation}
At $n_f=n_f^c$ the fixed points merge and for $n_f<n_f^c$ the beta function looses the non-trivial fixed points. For negative $\delta$ within the following range: 
\begin{equation}
-\frac{1}{8} < \delta \leq 0
\end{equation}
there is a global maximum of the beta function at the origin, a local minimum at $\alpha=\frac{1}{4}(3-\sqrt{1+8\delta})$ and a local maximum at $\alpha=\frac{1}{4}(3+\sqrt{1+8\delta})$. For illustration we plot the beta function for different values of $\delta$ in Fig.~\ref{Beta_Miransky}.
It is possible to find an analytical solution to the RG equation: 
\begin{equation}
{d \ln \mu} =  \frac{d\alpha}{\beta_{MY}} \ ,
\end{equation}
reading: 
\begin{equation}
\hskip -.8cm 
\ln \frac{\mu}{\mu_0} = \frac{\alpha  (1+\delta ) \text{ArcTanh}\left[\frac{-1+\alpha }{\sqrt{\delta }}\right]+\sqrt{\delta } \left(1-\delta  +\alpha  \ln\left[\frac{(\alpha -1)^2-\delta}{\alpha^2}\right]\right)}{\alpha  (-1+\delta )^2 \sqrt{\delta }}\Big|^{\alpha(\mu)}_{\alpha(\mu_0)} \ .
\end{equation}
If we were to consider the case of $\delta$ positive, but smaller than unity so that asymptotic freedom is kept, we would discover that there are three distinct branches. The one to the left of the IR fixed point, the one where $\alpha$ is in between the nontrivial IR and UV fixed point, and the one to the right of the nontrivial UV fixed point.  To the left of the IR fixed point one starts the flow from any $\mu_0$ sufficiently close to the trivial UV fixed point and one ends up at the attractive IR fixed point. Another {\it asymptotically safe} theory is defined in between the two non-trivial fixed points. In this region the coupling runs at low energies to the IR fixed point and raises at high energies till the non-trivial UV fixed point is reached. Finally, to the right of the non-trivial UV fixed point the theory, and hence beta function, runs in the deep infrared to increasingly large values of the coupling.

We turn now our attention to negative values of $\delta$ corresponding to the phase in which the beta function features no non-trivial fixed points. To elucidate the near-conformal dynamics  we investigate the region $ - 1/8< \delta < 0$. In particular we will consider the limit $-\delta = n_f^c - n_f \rightarrow 0$. In the deep infrared the coupling constant runs to infinity and we start the running in the UV near $\alpha =1$ at $\mu_0$. With these boundary conditions we find: 
\begin{equation}
\Lambda_{MY} =\frac{\mu_0}{n_f^c - n_f} \exp\left[ - \frac{\pi}{2\sqrt{n_f^c - n_f}}\right]  \ ,~ n_f \rightarrow n_f^c\ , ~  {n_f \leq n_f^c} \ . 
\label{MYScale}
\end{equation}
Here $\Lambda_{MY}$ is the infrared scale to be identified, for example, with a physical scale of the theory such as the mass of a hadron. This scale vanishes exponentially fast when approaching the critical number of flavors above which the infrared fixed point is generated. This exponential behavior is the essence of the Miransky scaling.  

\begin{figure}[h!]
  ~\vfill
 \begin{center}
\includegraphics[width=8truecm]{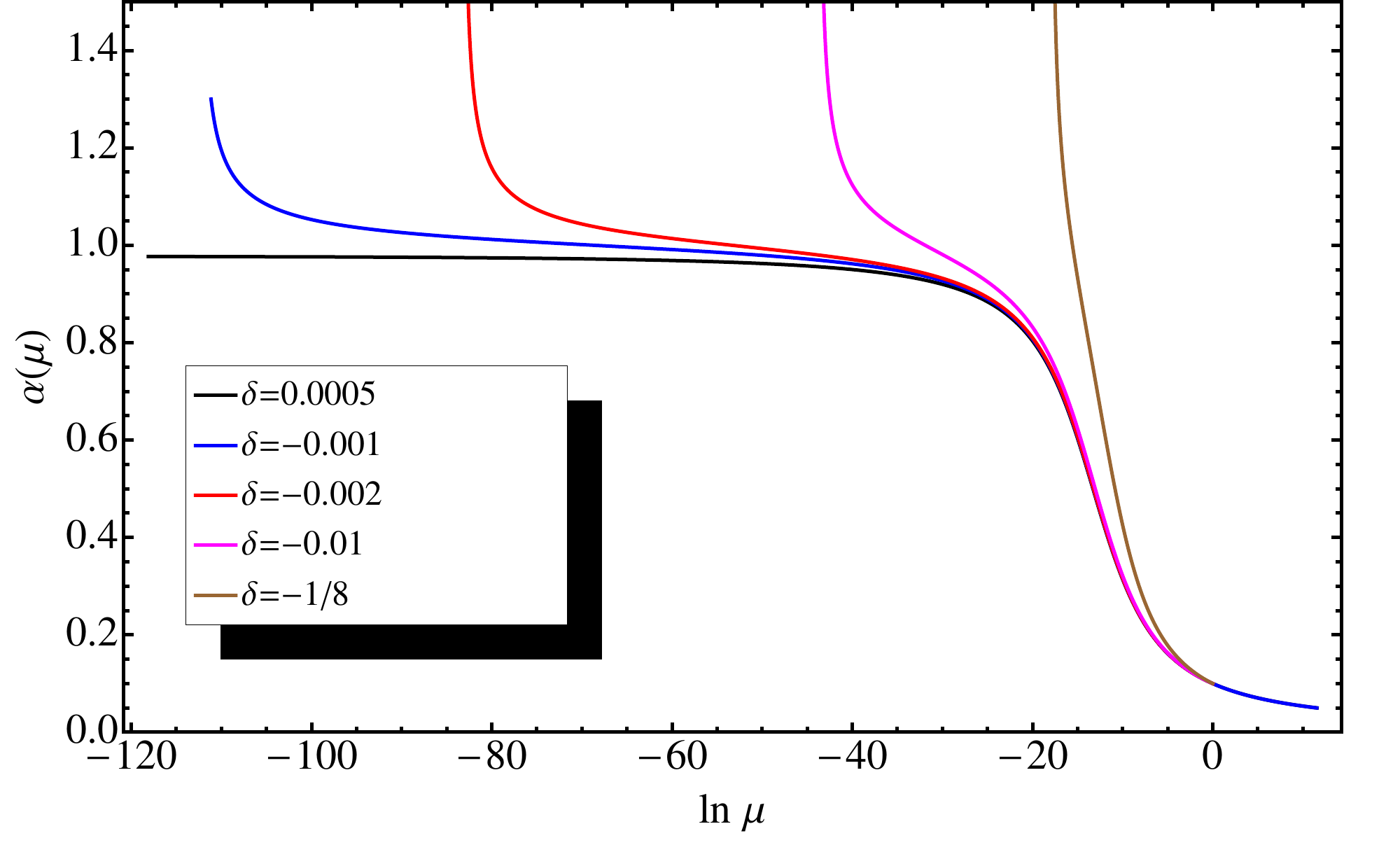}
\caption{Running of the coupling constant coming from $\beta_{MY}$ for different values of $\delta = n_f - n_f^c$ within the {\it walking} regime. All the solutions are normalized at $\mu_0$ such that $\alpha(\mu_0)=0.1$ and $\mu$ is normalized to $\mu_0$. } \label{running_miransky}
\end{center}
\end{figure}

In Fig.~\ref{running_miransky} we plot the running of the coupling constant for different negative values of $\delta$ with the normalization condition $\alpha(\mu_0)=0.1$. The figure visualizes the idea of walking dynamics introduced by Holdom \cite{Holdom:1981rm,Holdom:1984sk} and further crystallized in \cite{Yamawaki:1985zg,Appelquist:1986an}. In lay terms the coupling constant runs slowly, i.e. {\it walks}, towards the infrared value remaining near constant over a range of energies becoming wider and wider as one approaches, as function of $\delta$, the double fixed point. 
Further assuming that the beta function corresponds to an underlying gauge theory featuring fermions we now determine the scaling behavior of the chiral condensate of the theory in the walking regime. Defining with $\gamma$ the anomalous dimension of the mass of the Dirac fermion $Q$ in a given representation of an underlying gauge group we have the following well-known RG equation: 
\begin{eqnarray} \langle\bar QQ\rangle_\mu &=&
\exp\left(\int_{\Lambda}^{\mu}
\text{d}(\ln\mu)\gamma(\alpha(\mu))\right)\langle\bar QQ\rangle_\Lambda \nonumber \\ &=&\exp\left(\int_{\alpha({\Lambda})}^{\alpha(\mu)}
\text{d}\alpha\frac{\gamma(\alpha)}{\beta(\alpha)}\right)\langle\bar QQ\rangle_\Lambda \ 
\label{condensate}
\end{eqnarray}
relating the condensate at two different energies. Using $\beta_{MY}$ case in the walking region we have 
\begin{widetext}
\beq \langle\bar QQ\rangle_\mu =
\exp\left(\int_{\alpha({\Lambda})}^{\alpha(\mu)}
\text{d}\alpha\frac{\gamma(\alpha)}{-\alpha^2 ((\alpha -1)^2 + |\delta|)}\right)\langle\bar QQ\rangle_\Lambda \ \simeq 
\exp\left(\gamma (1) \int_{\alpha({\Lambda})}^{\alpha(\mu)}
\text{d}\alpha\frac{1}{\beta_{MY}}\right)\langle\bar QQ\rangle_\Lambda \ = \left(\frac{\mu}{\Lambda}\right)^{\gamma(1)}  \langle\bar QQ\rangle_\Lambda \ .
\label{condensateMY}
\eeq 
\end{widetext}
In the last passage we have used the definition of the beta function and, in the first step, assumed that the anomalous dimension of the mass operator is smooth across the phase transition. $\gamma(1) = \gamma(\alpha_{IR}=\alpha_{UV})$ is the value of the anomalous dimension at the merger. We have re-derived the power-law enhancement of the chiral condensate with the energy distinctive of walking dynamics. Since $\gamma$ is evaluated at the fixed point its value is scheme-independent \cite{Pica:2010mt}.  If we further model $\gamma = \alpha$ we have
$\gamma({\alpha_{IR/UV}}) = 1 \mp \sqrt{\delta}$ , with $\gamma({\alpha_{IR}}) + \gamma({\alpha_{UV}}) = 2$.
\section*{Jumping Dynamics}
\begin{figure*}[ht!]
	{  \includegraphics[width=0.44\textwidth]{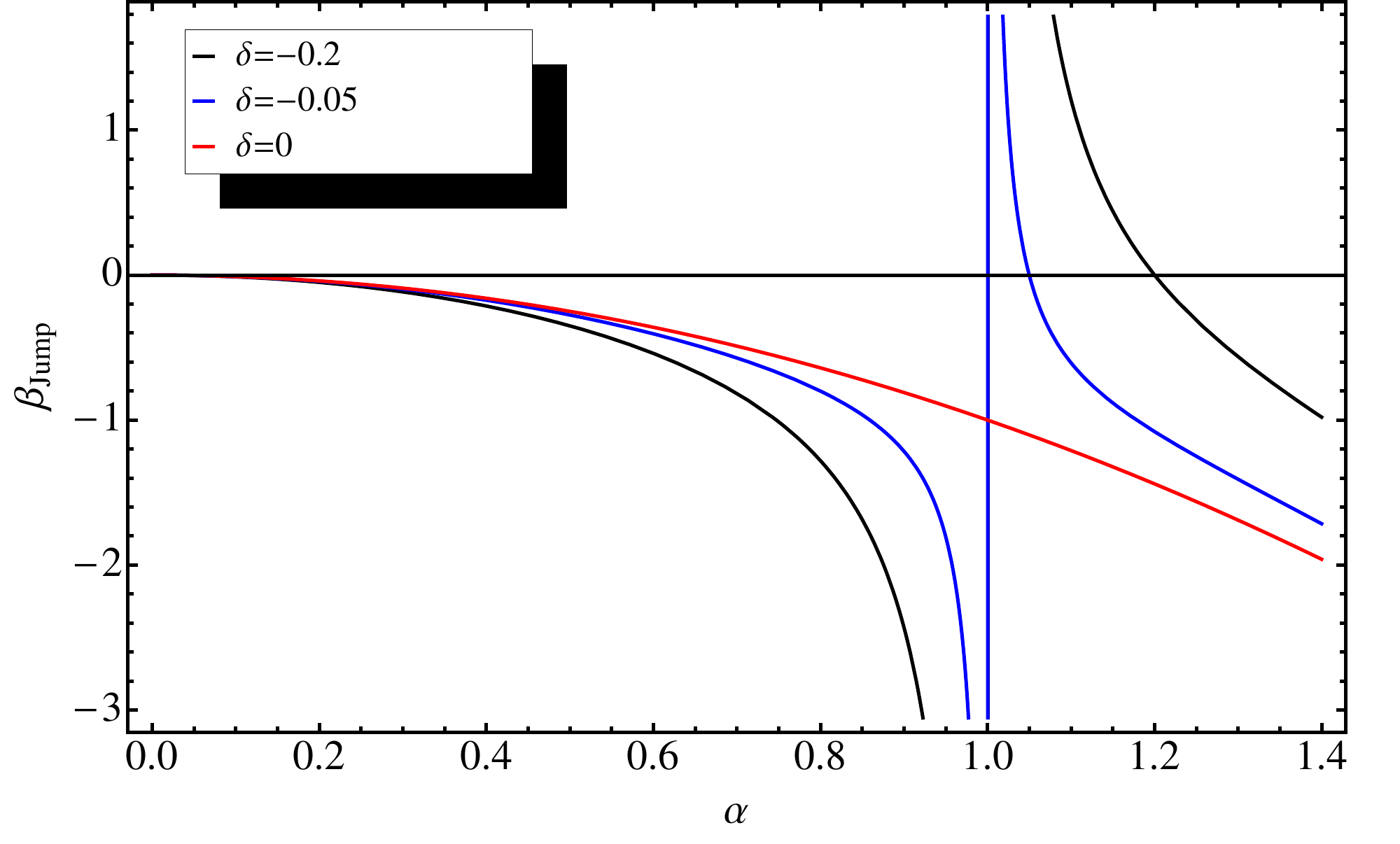}}
	\hfill
	{ \includegraphics[width=0.44\textwidth]{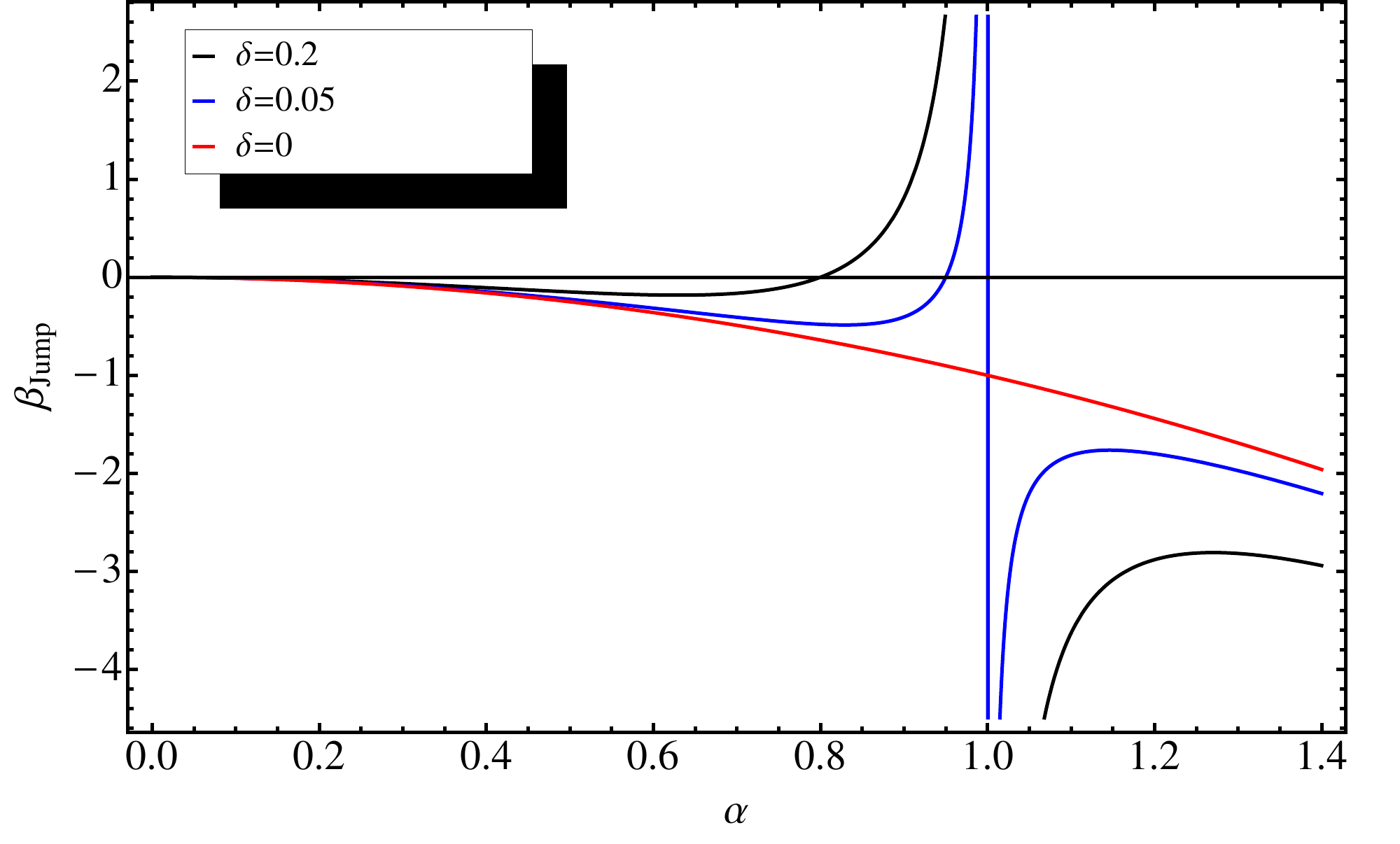}} 
		\caption{$\beta_{Jump}$ for different values of $\delta = n_f - n_f^c$.} \label{betajump}
\end{figure*}

The previous section embodies the standard paradigm of walking dynamics. However this picture is far from established analytically or via first principle lattice simulations. It is therefore relevant to consider other theoretical scenarios and their impact on particle physics phenomenology. We start by observing that there is the logical possibility that the full beta function of the theory develops, at least, a zero in the denominator. This occurs exactly for supersymmetric gauge theories \cite{Novikov:1983uc} and the all-orders beta function conjectured to be valid also for non-supesymmetric gauge theories with fermionic matter \cite{Pica:2010mt,Ryttov:2007cx}. Moreover it is reasonable to expect that the full perturbative and non-perturbative contributions to the beta function conspire to generate a non-trivial pole structure \cite{Chishtie:1999tx}. Whatever the pole structure is, if the underlying theory displays conformality, there will be also zeros in the numerator of the beta function associated to the non-trivial fixed point structure of the theory. Here we consider the simplest example in which the beta function has a simple nontrivial zero in the numerator and a simple pole. We will always assume the existence of the trivial double zero at the origin so that the beta function contains information about the asymptotically free nature of the theory. Without loss of generality we write: 
\begin{equation}
\beta_{Jump} = - \alpha^2 \frac{1-\delta - \alpha}{1-\alpha} \ .
\end{equation}
By construction this beta function has a zero in the numerator for any $\delta$ 
which is to the left of the pole value $\alpha_{pole} =1$ for $\delta$ positive and to the right for $\delta$ negative. Here we take again $\delta = n_f - n_f^c$. It is straightforward to show that this zero corresponds to an IR (UV) fixed point for $\delta>0$ ($\delta <0$) : 
\begin{equation}
\alpha_{IR(UV)} =  1 - \delta \ , \quad \delta >0 ~~~~(\delta < 0) \quad {\rm and} \quad | \delta| \leq 1\ .
\end{equation} 
Because of the presence of the pole the beta function describes two disconnected theories. One which is continuously connected to the asymptotically free underlying gauge theory and the other which is not.
We plot the beta function for positive and negative values of delta in Fig.~\ref{betajump}. At exactly $\delta=0$ the numerator and denominator of the beta function cancel and we are left with $\beta_{Jump}^{\delta=0}= -\alpha^2$ which is the red curve in Fig.~\ref{betajump}. 
  What happens at the phase boundary? We will demonstrate that there is a sudden jump as we drop the number of flavors below the critical number of flavors (i.e. at $\delta=0$) of the intrinsic physical scale of the theory. 
 
\begin{figure} [bh!]
\begin{center}
\includegraphics[width=8truecm]{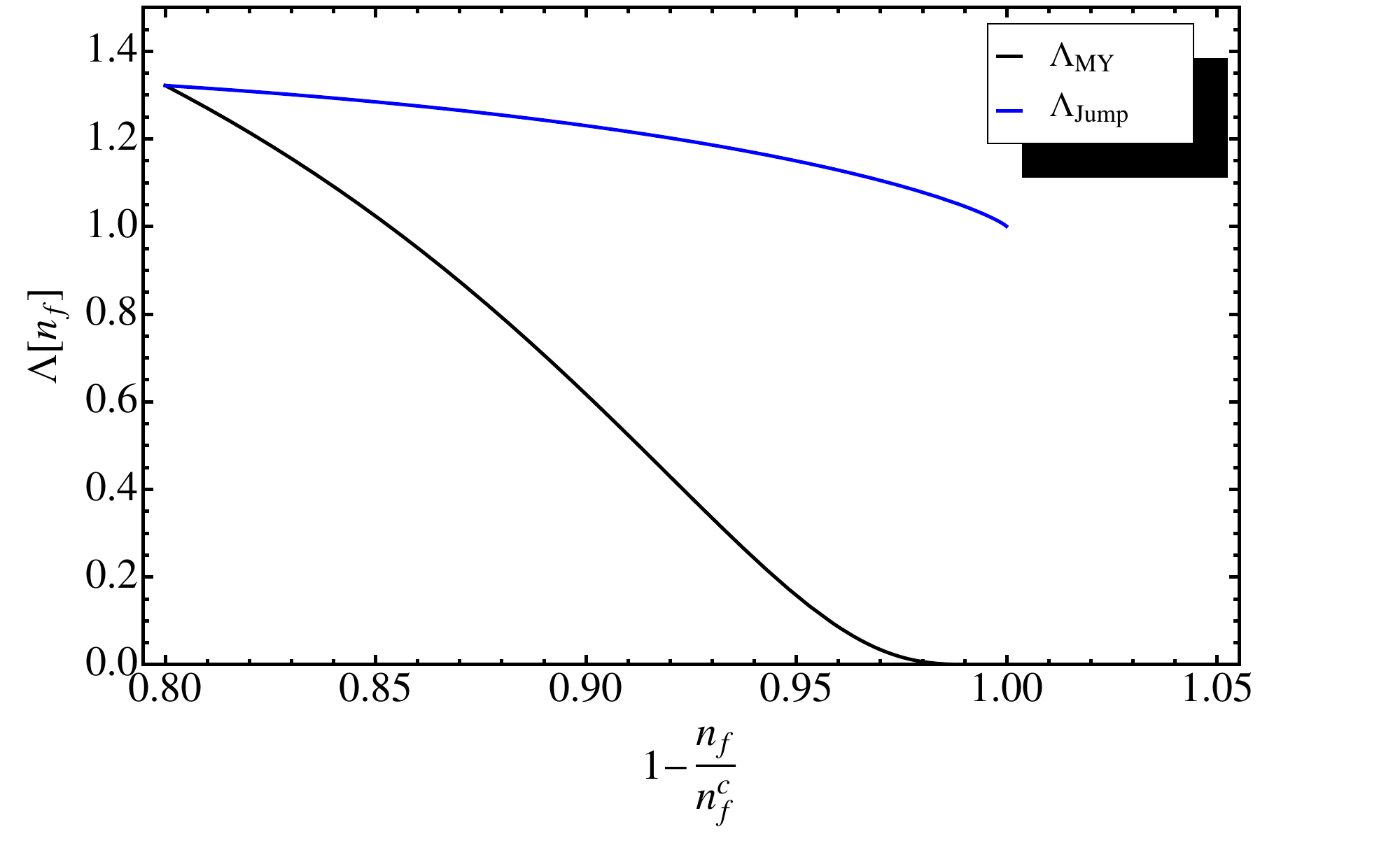}
\caption{Mass gap dependence on the number of flavors for the Miransky scaling (lower black curve) and for the jumping dynamics (upper blue curve). } \label{jump}
\end{center}
\end{figure}

 We start by constructing the analytical solution for the RG equation of the coupling which reads: 
 \begin{equation}
 \ln \frac{\mu}{\mu_0} = \frac{1}{\alpha(1-\delta)}\Big|^{\alpha(\mu)}_{\alpha(\mu_0)} + \frac{\delta}{(1-\delta)^2} \ln\left[\frac{1-\delta- \alpha}{\alpha}\right]\Big|^{\alpha(\mu)}_{\alpha(\mu_0)} \ .
 \end{equation}
 Holding fixed, as done for the Miransky scaling case, the coupling constant at a given renormalization scale one observes that the newly generated scale increases with decreasing the number of flavors below the conformal window in the following way:
 \begin{equation}
 \Lambda_{Jump}= \Lambda_{c} \left[ 1 - (n_f^c - n_f) \ln\left(n_f^c - n_f \right)\right]\ , ~ n_f \rightarrow n_f^c \ , ~  {n_f \leq  n_f^c} \ . 
 \end{equation}
  \begin{figure}[bh!]
\begin{center}
\includegraphics[width=8truecm]{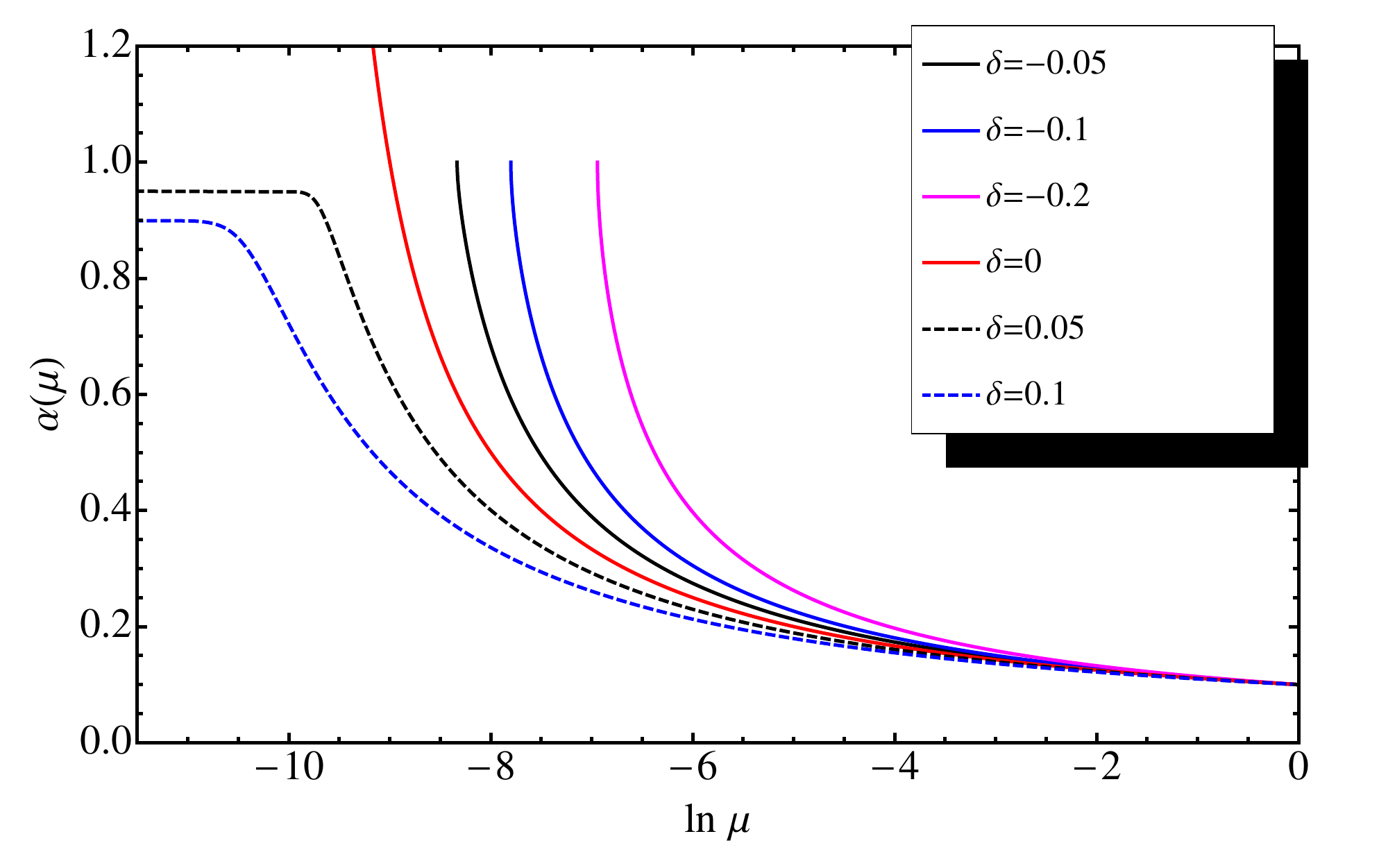}
\caption{Running of the coupling constant due to $\beta_{Jump}$ for different values of $\delta = n_f - n_f^c$. All the solutions are normalized at $\mu_0$ such that $\alpha(\mu_0)=0.1$, and of course $\mu$ is normalized to $\mu_0$. } \label{running_jump}
\end{center}
\end{figure}
  $\Lambda_c = \mu_0 \exp\left[ \frac{\ln \alpha_0}{\alpha_0} \right] $ is the renormalization group invariant scale of the theory at the critical number of flavors. However for $n_f>n_f^c$ no infrared scale is generated and necessarily there must be a {\it jump} in the spectrum from $\Lambda_c$ to zero. This result shows that $\beta_{MY}$ and $\beta_{Jump}$ describe two distinct physical systems. For illustration we summarize in Fig.~\ref{jump} the behavior of the physical scale of the theory, as function of number of flavors, for Miransky scaling and jumping dynamics.  To compare the two scaling laws we normalized the two scales at a given value of $n_f$. 

{\it Can jumping dynamics be used for models of dynamical electroweak symmetry breaking?}  At the conformal boundary the dynamics is QCD like and therefore one observes only a logarithmic enhancement of the condensate of the type $ \langle\bar QQ\rangle_\mu \simeq
\gamma({1})~\ln \left(\frac{\mu}{\Lambda}\right) \langle ~\bar QQ\rangle_\Lambda
$. The jumping dynamics does not lead to power-law enhancement of the chiral condensate required for {\it walking} Technicolor. Furthermore the $S$-parameter in the jumping scenario automatically respects the lower bound put forward in \cite{Sannino:2010ca} given that, opportunely normalized, at the phase boundary is as small as the one for QCD. Henceforth the answer to the original question is that one can break the electroweak theory via jumping dynamics but cannot accommodate the generation of the standard model fermion masses following the walking paradigm nor drastically reduce the QCD-like $S$-parameter.

We have shown that it is possible to devise a simple framework according to which the approach to the long distance conformality does not display any sign of walking dynamics.  All the current lattice investigations of the conformal window are not able to differentiate walking from jumping. The reasons being that: i) The lattice results, for the moment, are performed for a fixed number of flavors and therefore either there is a nonzero infrared scale or the theory is conformal; ii)  The precise determination of the chiral condensate is not a simple task making harder to disentangle a power-law from a logarithmic enhancement of the condensate as function of the renormalization scale as well as the number of flavors; iii) Even if the underlying dynamics is of walking type (with or without the introduction of four-fermion interactions) the extension of the region in the number of flavors and four-fermion coupling is not known and might be tiny; iv) Measuring a large anomalous dimension of the mass is encouraging but alone insufficient to demonstrate the existence of walking dynamics.

Because of the discontinuity of the order parameter at the conformal phase transition, i.e. of the vacuum expectation value of the trace of the improved energy momentum tensor which is proportional to the intrinsic scale of the theory, jumping dynamics corresponds to a first order {conformal phase transition}. First order phase transitions are common in nature  and therefore we expect jumping dynamics to constitute a likely scenario with inevitable important consequences on a large number of research fields ranging from a better understanding of strong dynamics and its holographic engineering to the construction of sensible extensions of the standard model. 

I thank O. Antipin, A. Hietanen, M. Mojaza, C. Pica, T. Ryttov, J. Schechter and U. Sondegaard  for useful discussions and careful reading of the manuscript. 

\end{document}